\documentstyle[preprint,aps]{revtex}
\draft
\begin{document}
\title{On mixing angles and resonances in three neutrino oscillations
in matter}
 \author{ V.M. Aquino}
\address{  Departamento de F\'\i sica \\ Universidade Estadual de
Londrina, UEL\\
 86000-970 Londrina, Paran\'a, Brasil}
\author {J. Bellandi and M.M. Guzzo}
\address{ Instituto de F\'\i sica `Gleb
Wataghin' \\  Universidade Estadual de Campinas, Unicamp \\
13083-970 Campinas, S\~ao Paulo, Brasil  }
\maketitle
\begin{abstract}
We find exact analytical expressions for  mixing angles in matter in
the context of three generation neutrino oscillations in matter to
discuss the role of resonances in this phenomenon. We show that some
knowledge from conventional two neutrino  MSW effect, which has been
extended to approximated solutions to three neutrino oscillations, has
to be abandoned in this exact approach. We observe that maximal values
for the mixing angles in matter are found in nonresonant regions  and
stationary phases do not coincide anymore with resonances in this
simple extension of the MSW effect. We present a general way to
identify a resonance and discuss what we can physically expect in these
regions.

\end{abstract}

\date{\today}

\newpage

\section{Introduction}

Resonant regions are previleged zones for neutrino conversion.
Concerning solar neutrinos, the importance of a resonance can be
appreciated remenbering that the standard MSW solution to the solar
neutrino problem requires values for the mixing angle
in vacuum $\theta$ and for the squared mass  difference $\Delta =
m^2_2-m^2_1$ such that $\sin^2 2 \theta < 10^{-3}$ and $\Delta \sin^2
2\theta \approx 10^{-8}$ eV$^{2}$ \cite{Barger91} which imply a
resonance in the neutrino trajectory inside the sun when the
approximately exponentially decreasing standard solar matter
distribution is assumed \cite{Bahcall}. This is the so-called
nonadiabatic solution to the solar neutrino problem and the role of the
resonance is evident in such situation once that it is well known that
the adiabaticity parameter \cite{MS} presents its smallest
values  in a resonance region, which imply that neutrino transitions
are less adiabatic in that region.

Resonances in two family MSW effect \cite{MS,W} are associated with maximum
mixing
between the two flavor eigenstates. This can be appreciated investigating
the behavior of the matter mixing angle when the relevant matter density
varies along the neutrino trajectory. The mixing angle in matter $
\tilde\theta$ is introduced as the parameter that characterizes a rotation
of the two-dimensional neutrino space from the basis of the current
eigenstates $(\nu_e,\nu_\mu)$ to the basis of the physical eigenstates
$(\nu_1,\nu_2)$:

\begin{equation}
\label{eq1}
\begin{array}{c}
\nu_1(t)=\nu_e(t)\cos\tilde\theta(t) -
\nu_\mu(t)\sin\tilde\theta(t), \\
\nu_2(t)=\nu_e(t)\sin\tilde\theta(t) +
\nu_\mu(t)\cos\tilde\theta(t).
\end{array}
\end{equation}

\noindent It can be calculated \cite{BP}:

\begin{equation}
\label{eq2}
\sin ^22\tilde \theta (t)=\frac{\sin ^22\theta }
{\left[ \frac{ 2E\sqrt{2}G_FN_e(t)}{\Delta} -\cos 2\theta \right]^2
+\sin^22\theta },
\end{equation}
where $E$ is the neutrino energy and $G_FN_e(t)$ is the consequence of
electron neutrino coherent forward scattering from electrons in matter, the
number density of which at the region reached by neutrinos at instant $t$ is
$N_e(t)$.

{}From Eq.~(\ref{eq2}) it is possible to see that $\tilde\theta$ is
substantially modified by the neutrino coherent scattering from the medium.
If $N_e(t)\to 0$, $\tilde\theta\to \theta$ and we recover vacuum
expressions. When $N_e(t)$ is extremely large, $\tilde\theta\to \pi/2$ and $%
\nu_1\to -\nu_\mu$ while $\nu_2\to \nu_e$. An interesting intermediate case
occurs when

\begin{equation}
\label{eq3}N_e(t)=\frac{1}{2\sqrt{2}G_F} \frac{\Delta}{E} \cos 2\theta
\end{equation}
and the brackets in the denominator of Eq. (\ref{eq2}) vanishes. In this
point the mixing of flavor eigenstates is maximal, i.e., from Eq. (\ref{eq1})
we see that the probability of finding an electron or a muon neutrino in any
of the mass eigenstates is 1/2. This feature has been used to characterize a
resonance: the maximum of the bell-shaped $\sin^2 2 \tilde \theta \times N_e$
graph indicates a resonance.

The resonance condition given by Eq. (\ref{eq3}) coincides also with the
position where the difference of the two squared matter eigenvalues of the
corresponding time evolution matrix in matter $\tilde m_2^2-\tilde  m_1^2$
is a minimum, suggesting
that the resonance is the region where transitions between matter
eigenstates are most likely to happen.

Finally, it was noticed in reference \cite{GB92} that the resonance
condition (\ref{eq3}) coincides also with the condition of existence of a
stationary phase \cite{erdely} in the two neutrino time evolution equations.
Such fact
allows to investigate the evolution of this neutrino system around a
resonance calculating, through the stationary phase method \cite{erdely}, the
related Green function. Employing this method it was possible to evaluate
\cite{GBA93}
the level crossing probability, i.e., the probability of nonadiabatic
transitions between matter eigenstates $\nu_1$ and $\nu_2$ as an alternative
approach to Landau-Zener \cite{LZ} or Petcov \cite{Petcov87} methods.

In this paper we investigate how is the behavior of mixing angles in
matter and how to identify a resonance in the context of a three
neutrino system oscillating in matter. We assume standard electroweak
interactions of neutrinos with matter as well as nonvanishing vacuum
mixing angles and nondegenerated mass eigenstates (in vacuum).
Therefore we are analysing the simplest extension of the conventional
MSW effect \cite{MS,W} to the case where three families are
present. We verify that the above mentioned three criteria usually used
to define a resonance in two neutrino matter oscillations, namely,
maximal mixing angles in matter, minimal eigenvalue difference and the
presence of a stationary phase, do not lead anymore to the same region
in the neutrino trajectory. Note also that these same criteria have
been used in approximated solutions to three neutrino oscillations in
matter \cite{KP,Barger}. Consequently some of them have to be
abandoned. We present, therefore, based on exact analytical expressions
for mixing angles in matter, how we can use our previous knowledge
coming from two neutrino matter oscillations to arrive to a solid
condition defining resonances in three neutrino oscillations and,
therefore, an accurate analytical description of the physical
consequences around such regions.

\section{ANALYTICAL SOLUTION}

A general time evolution equation describing a three level system can be
written as an equation for a three-component spinor $\Phi (t)\equiv (\Phi
_1,\Phi _2,\Phi _3)$:
\begin{equation}
\label{Eq.1}
i\frac d{dt}\Phi (t)=h(t)\Phi (t),
\end{equation}
where  the hamiltonian $h(t)$ is a
$3\times 3$ matrix which elements are specified according to the dynamical
situation from which a  boundary condition $\Phi (t_o)$ is given.
A general solution of Eq. (\ref{Eq.1}) can be written in the the form
\begin{equation}
\label{Eq.2}\Phi (t)=Exp \left[ -i\int_{t_0}^t h(t^{\prime })dt^{\prime }
\right]\Phi (t_o),
\end{equation}
where the symbol $Exp$ represents a sum of multiple time
ordered integrals \cite{Feynman}.

For a time-independent hamiltonian, the solution of Eq. (\ref{Eq.1})
can be obtained by means of the Laplace transformation. Introducing
the Laplace transformed $\Psi (p)=L \left[\Phi (t)\right]$, then
\begin{equation}
\label{Eq.3}
p\Psi (p)-\Phi (t_o)=-ih\Psi (p)
\end{equation}
and
\begin{equation}
\label{Eq.4}
\Phi (t)=L^{-1}\left[ (p{\bf 1}+ih) ^{-1}\right]\Phi (t_o).
\end{equation}
 The solution $\Phi (t)$ depends on the elements of the $h$ matrix and on the
roots $\lambda _i$ $(i\equiv 1,2,3)$ of the characteristic polynomial of the
$h$ matrix
\begin{equation}
\label{Eq.5}
\det \left[ p{\bf 1}+ih \right] =0.
\end{equation}

In the particular case we are interested in, where a three neutrino system
oscillates in matter, interacting with it through standard electroweak
interactions,
the $h$ matrix is given by
\begin{equation}
\label{Eq.6}
h=\frac 1{2E}\left[  UM^2U ^{-1}+A\right],
\end{equation}
where $M^2$ is a diagonal matrix given by
\begin{equation}
\label{Eq.7}(M^2)_{ij}=m_i^2\delta _{ij},
\end{equation}
$m_i^2$ are the three neutrino squared mass eigenvalues in  vacuum,
\begin{equation}
\label{Eq.8}U=e ^{i\psi \Lambda _7}\Gamma e ^{i\phi \Lambda _5}e ^{i\omega
\Lambda _2}
\end{equation}
is the $3\times 3$ mixing matrix where
$\Lambda _i$ are the Gell-Mann matrices, $\psi ,\phi $ and $
\omega $ are the mixing angles in  vacuum and $\Gamma $ is a matrix
containing complex phases that we will ignore since we assume CP
conservation $(\Gamma \equiv 1).$

Since we  consider here only  standard neutrino interactions with ordinary
matter, $A$ matrix has its first element  $A_{11}$ given by
\begin{equation}
\label{Eq.9}A_{11}=2\sqrt{2}G_FN_eE
\end{equation}
and all others are zero. Note that neutral current contributions to $A$
are proportional to the unit matrix, giving only irrelevant overall phases
to the final solution of Eq. (\ref{Eq.1}). $G_F$, $E$ and $N_e$ were previously
introduced.

For neutrino propagating in vacuum,  $A=0$, and the solution of Eq.
(\ref{Eq.1})
is trivial and simply given by
\begin{equation}
\label{Eq.10}
\Phi (t)=Um ^2U ^{-1}\Phi (t_o),
\end{equation}
where $m ^2$ is a diagonal matrix with elements
\begin{equation}
\label{Eq.11}(m ^2)_{ij}=\exp \left[ -i\frac t{2E}m_i^2 \right]\delta _{ij}.
\end{equation}

The solution of Eq. (\ref{Eq.1}) in matter, with $A$ being a time-dependent
matrix,
is given by Eq. (\ref{Eq.2}) and it depends on the specific $N_e$ function
describing
the electron density. However, when $A$ can be considered a constant matrix,
as it is  supposed in the adiabatic approximation, Eq. (\ref{Eq.1}) has an
exact
 analytical
solution, obtained by Laplace transformation. Furthermore, the $A$ matrix is
invariant under a $e ^{i\psi \Lambda _7}$ rotation, then, introducing now
\begin{equation}
\label{Eq.11a}\Psi (t)=e  ^{-i\psi \Lambda _7}\Phi (t),
\end{equation}
we observe that $\Psi (t)$ satisfies the following differential equation
\begin{equation}
\label{Eq.12}\frac d{dt}\Psi (t)=-iH\Psi (t),
\end{equation}
with boundary condition $\Psi (t_o)=e ^{-i\psi \Lambda _7}\Phi (t_o)$ and

\begin{equation}
\label{eq.13}H=\frac 1{2E} \left[ e ^{i\phi \Lambda _5}e ^{i\omega \Lambda
_2}M ^2e ^{-i\phi \Lambda _5}e ^{-i\omega \Lambda _2}+A\ \right],
\end{equation}
which can be explicitly written as
\begin{equation}
\label{eq.14}{H=\frac 1{4E}\left(
\begin{array}{ccc}
\Lambda \cos {} ^2\phi +2m_3 ^2\sin {} ^2\phi +2A & \Delta \sin 2\omega \cos
\phi & (m_3 ^2-\frac \Lambda 2)\sin 2\phi \\
\Delta \sin 2\omega \cos \phi & \Sigma +\Delta \cos 2\omega & -\Delta \sin
2\omega \sin \phi \\
(m_3 ^2-\frac \Lambda 2)\sin 2\phi & -\Delta \sin 2\omega \sin \phi & \Lambda
\sin ^2\phi +2m_3 ^2\cos ^2\phi
\end{array}
\right) }
\end{equation}
where $\Delta =m_2 ^2-m_1 ^2$, $\Delta _1=m_2 ^2+m_1 ^2-2m_3 ^2$, $\Sigma =m_2
^2+m_1 ^2$
and $\Lambda =\Sigma -\Delta \cos 2\omega $.

On the Laplace space we have
\begin{equation}
\label{Eq.15}
\Psi (p)= \left[ p{\bf 1}+iH\right] ^{-1} \Psi (t_o).
\end{equation}

To calculate $\Psi (t)$ we have to obtain the roots of the characteristic
polynomial of the matrix $H$, $\det \left[ p{\bf 1}+iH \right] =0,$ which are
given by \cite{Bateman}
\begin{equation}
\label{Eq.16}\lambda _1=\frac{m_1 ^2+m_2 ^2+m_3 ^2+A}{6E}-\frac
1E\sqrt{\frac{-Q
}3}\cos \frac \alpha 3,
\end{equation}
\begin{equation}
\label{Eq.17}\lambda _2=\frac{m_1 ^2+m_2 ^2+m_3 ^2+A}{6E}+\frac 1{2E}\sqrt{
\frac{-Q}3}\cos \frac \alpha 3-\frac 1{2E}\sqrt{-Q}\sin \frac \alpha 3,
\end{equation}
\begin{equation}
\label{Eq.18}\lambda _3=\frac{m_1 ^2+m_2 ^2+m_3 ^2+A}{6E}+\frac 1{2E}\sqrt{
\frac{-Q}3}\cos \frac \alpha 3+\frac 1{2E}\sqrt{-Q}\sin \frac \alpha 3,
\end{equation}
where
\begin{equation}
\label{Eq.19}
\alpha =\arccos \frac{-R}{2\sqrt{\frac{-Q ^3}{27}}},
\end{equation}
\begin{equation}
\label{Eq.20}Q=\frac{-1}{(2E) ^2}\left\{ \frac{\Delta  ^2}4+\frac{\Delta _1
^2}{%
12}+\frac{A ^2}3-\frac{A\Delta \cos {} ^2\phi \cos 2\omega }2+\frac{\Delta
_1A(\cos {} ^2\phi -2\sin ^2\phi )}6\right\}
\end{equation}
and
$$
R=-\frac{1}{27(2E) ^3}\left\{ \frac{\Delta _1 ^3}4-2A ^3-\frac{9\Delta
^2\Delta
_1}4+\frac{3\Delta _1 ^2A(\cos ^2\phi -2\sin ^2\phi )}4\right\} +
$$
$$
-\frac{1}{27(2E)^3}\left\{ \frac 92A ^2\Delta \cos  ^2\phi \cos 2\omega
-\frac 94A\Delta  ^2(\cos ^2\phi -2\sin ^2\phi )\right\} +
$$
\begin{equation}
\label{eq.21}
-\frac{1}{27(2E)^3}\left\{ \frac 92\Delta _1A\Delta \cos
^2\phi \cos 2\omega -\frac 32A^2\Delta _1(\cos ^2\phi -2\sin ^2\phi
)\right\}.
\end{equation}

Note now that in vacuum we have

\begin{equation}
\lambda _1 ^v=\frac{m_3 ^2}{2E},~~\lambda _2 ^v=\frac{m_1 ^2}{2E}~~{\rm  and}~~
\lambda _3 ^v=\frac{m_2 ^2}{2E},
\end{equation}
 and the Laplace anti-transformation of Eq. (\ref{Eq.15}) reproduces the
corresponding solution given by Eq. (\ref{Eq.10}).
 The roots $\lambda _i$ of the characteristic polynomial are the squared mass
eigenvalues in matter. Because of the arbitrariness in the choice of the
order of the roots, we use the above vacuum limit to order the roots in terms
of
the squared mass eigenvalues in the matter. We define:
\begin{equation}
\lambda _1=\frac{\tilde  m_3^2}{2E},~~\lambda _2=\frac{\tilde m_1 ^2}{2E}~~{\rm
and} ~~
\lambda _3=\frac{\tilde m_2 ^2}{2E}.
\end{equation}

Finally, we can write the solution of Eq. (\ref{Eq.1}) in terms of a $T$
transition
 matrix such that
\begin{equation}
\label{Eq.22}\Phi (t)=e ^{i\psi \Lambda _7}Te ^{-i\psi \Lambda _7}\Phi (t_0),
\end{equation}
where the elements of the $T$ matrix, given in terms of the $\lambda _i$
roots and of the elements of the $H$ matrix, can be written as:

$i)$ diagonal elements:
\begin{equation}
\label{Eq.23}T_{ii}=\sum_{m=1} ^3C_m\left[  (\lambda _m-H_{jj})(\lambda
_m-H_{kk})-H_{jk} ^2\right] e^{-i\lambda _mt},
\end{equation}

$ii)$ non diagonal elements $(T_{ij}=T_{ji})$:
\begin{equation}
\label{Eq.24}T_{ij}=\sum_{m=1} ^3C_m\left[  H_{ij}(\lambda
_m-H_{kk})-H_{ik}H_{jk}\right] e ^{-i\lambda _mt},
\end{equation}
where
\begin{equation}
\label{Eq.25}C_m=\left[ (\lambda _m-\lambda _\ell )(\lambda _m-\lambda
_n)\right]  ^{-1}
\end{equation}
with $m\neq \ell \neq n$ and $n,\ell ,n\equiv (1,2,3).$

Note also that all well known results for a two neutrino system oscillating
in matter can be
straightforwardly obtained from the solution given by Eq. (\ref{Eq.22}).

\section{MIXING ANGLES IN MATTER}

It is well known that the knowledge of the mixing angles in  matter is
important to study resonant transitions between flavor neutrino states
\cite{BP}. In order to explicitly write an exact expression for these
angles, we  define $\tilde \psi ,\tilde \phi $ and $\tilde \omega $ as
the mixing angles in the matter. We can write therefore the final
solution of Eq. (\ref{Eq.1}) in terms of mixing angles in  matter in
analogy with what  we did in the vacuum case, Eq.(\ref{Eq.10}), using
now the final solution given by Eq. (\ref{Eq.22}).
This solution can be written in the following
way
\begin{equation}
\label{Eq.27}\Phi (t)=U(\tilde \psi ,\tilde \phi ,\tilde \omega
)M_m ^2U ^{-1}(\tilde \psi ,\tilde \phi ,\tilde \omega )\Phi (t_0)
\end{equation}
where
\begin{equation}
\label{Eq.28}(M_m ^2)_{ij}=\exp [ -it\lambda _i]\delta _{ij}
\end{equation}
and
\begin{equation}
\label{Eq.29}
U(\tilde \psi ,\tilde \phi ,\tilde \omega )=e ^{i\tilde \psi
\Lambda _7}e ^{i\tilde \phi \Lambda _5}e ^{i\tilde \omega \Lambda _2}.
\end{equation}

In order to get the matter mixing angles we simply compare Eq. (\ref{Eq.27})
with Eq. (\ref{Eq.22}), and
after some algebra, we obtain
\begin{equation}
\label{Eq.30}\sin ^2\tilde \phi =\frac{\lambda _1 ^2-(H_{22}+H_{33})\lambda
_1+H_{22}H_{33}-H_{23} ^2}{(\lambda _1-\lambda _2)(\lambda _1-\lambda _3)},
\end{equation}
\begin{equation}
\label{Eq.31}\tan ^2\tilde \omega =\frac{ [\lambda
_3 ^2-(H_{22}+H_{33})\lambda _3+H_{22}H_{33}-H_{23}^2](\lambda _1-\lambda _2)
}{ [\lambda _2 2-(H_{22}+H_{33})\lambda _2+H_{22}H_{33}-H_{23}^2](\lambda
_3-\lambda _1)}
\end{equation}
and
\begin{equation}
\label{Eq.32}\tan {}\tilde \psi =\frac{(H_{12}\lambda
_1+H_{13}H_{23}-H_{12}H_{33})\cos \psi +(H_{13}\lambda
_1+H_{12}H_{23}-H_{22}H_{13})\sin \psi }{(H_{13}\lambda
_1+H_{12}H_{23}-H_{22}H_{13})\cos \psi -(H_{12}\lambda
_1+H_{13}H_{23}-H_{12}H_{33})\sin \psi }.
\end{equation}

\section{ANALYSIS OF THE RESULTS}

In Fig.~1 it is presented a comparison of the behavior of the quadratic
matter eigenvalues $\tilde m_i ^2$ and the relevant matter mixing angles $%
\tilde \omega$ and $\tilde \phi$ as a function of the parameter $A$ for
specially chosen values of vacuum parameters (see the corresponding caption
for details. $A$ is given in units of $m_1 ^2$). There are two resonances
clearly indicated by the minimum difference between the shown quadratic
masses. We observe that $\tilde \omega$ presents a maximum value in the
lower resonance (where $\tilde m ^2_2-\tilde m ^2_1$ is minimum) while $\tilde
\phi$ shows a maximum in the region of the higher resonance ($\tilde
m ^2_3-\tilde m ^2_2$ is a minimum). Interesting enough, differently from what
is expected in the two flavor neutrino oscillations in matter, the
conventional MSW effect, and also from what was found in previous
approximated analyses of the three neutrino oscillations \cite{KP,Barger},
a second peak for the mixing angle $\tilde \omega$ is found after the higher
resonance \cite{osc2}. In Fig.~2 we show the same graphs presented in Fig.~1 to
evidenced this unexpected behavior of the mixing angle $\tilde\omega$ for
larger values of $A$. It is clear from this figure that the criterion of
defining a resonance by means of localizing the maximal mixing angle in
matter, which can be safely used in two neutrino conventional MSW effect, leads
to
some ambiguity in the context of three neutrino oscillations and therefore
has to be abandoned.

Instead, we can improve this criterion analysing the content of Figs.~3 and
4. Note that the admixture of flavor eigenstates in each of the matter
eigenstates can be obtained through $\tilde \nu_i= \sum_\alpha
U_{i\alpha}\nu_\alpha$,
where $i=1,2,3$, $\alpha=e, \mu, \tau$ and $U_{i\alpha}$ is given by Eq.
(\ref{Eq.8}). Let us write now, as an example,  the linear combination of
flavor eigenstates in the first matter eigenstates:

\begin{equation}
\label{Eq.32a}\tilde\nu_1=\cos \tilde \phi \cos \tilde \omega \nu_e
+\cos\tilde\phi \sin\tilde\omega\nu_\mu +\sin\tilde\phi\nu_\tau .
\end{equation}

In Figs.~3 and 4 we show therefore the coefficients of this admixture
(values of the vacuum parameters are shown in the corresponding captions).
{}From Fig. 3 we observe that in the lower resonance the mixing of electronic
and muonic flavor eigenstates is maximal (when $\cos \tilde \phi \cos \tilde
\omega = \cos \tilde \phi \sin \tilde \omega$), while, from Fig.~4, we see that
the higher resonance coincides with the maximum admixture of $\nu_\mu$ and $%
\nu_\tau$, when  $\cos \tilde\phi\sin\tilde\omega = \sin\tilde\phi$.

Therefore, although we detected maxima of the mixing angles in matter in
regions far from resonances, it is still possible to identify a resonance
region searching for maximal mixing between flavor eigenstates. Note also
that such maximum are not anymore related with values of $\sqrt{2}/2$ for
flavor coefficients $|U_{i\alpha }|$ in the way it happened in the
conventional MSW phenomenon. This is because there could be nonnegligible
contributions from the flavor eigenstate that does not participate in the
resonant process. From the unitarity of the mixing matrix, we know that
$ \tilde U_{ie}| ^2+|\tilde U_{i\mu}| ^2+|\tilde U_{i\tau}| ^2=1$,
for $i=1,2,3$. Therefore, in the case where $i=1$ and $|\tilde U_{1\tau}|$ is
not vanishing, the maximal mixing is such that $|\tilde U_{1e}|=|\tilde
U_{1\mu}|<\sqrt{2}/2$. A similar situation occurs for the higher resonance
where we obtain $%
|\tilde U_{1\mu}|=|\tilde U_{1\tau}|<\sqrt{2}/2$. We can say that in three
neutrino oscillation phenomenon the mixing between flavor eigenstate around
a resonant region is as maximal as possible, although not in the same way as
in two neutrino oscillations, where maximum mixing implies that each one of
the neutrino flavor eigenstate participating in the resonant process
contributes with 50\% to the matter eigenstates.

A final issue to be discussed is the criterion of identifying a resonance
looking for a stationary phase in the neutrino evolution equations (\ref
{Eq.1}), in the same way it was proposed in reference \cite{GB92} in the
context of two neutrino MSW effect. A stationary phase is given by the
smallest difference of any two diagonal elements of the relevant evolution
matrix when one of these elements is time dependent. As an example, we quote
solar neutrinos where the matter density considerably varies along the
neutrino trajectory from the center of the sun, where neutrinos are created,
to the solar surface. Although in two neutrino oscillations this criterion
can be safely used, it does not work anymore in the presently analysed three
neutrino MSW effect. Stationary phases do not coincide with the minimum
squared mass differences or maximum flavor admixture.

Note however that it is still possible to use the stationary phase method to
calculate level crossing probabilities in the three neutrino oscillations.
Making convenient SU(3) transformations on the evolution matrix (\ref{Eq.6})
it is possible to conciliate resonances and stationary phases. This is
because resonances are invariant under similarity transformations, while
stationary phases do not. Therefore the matrix
\begin{equation}
\label{Eq.34}H ^1=e ^{-i\phi \Lambda _5}He ^{i\phi \Lambda _5}
\end{equation}
presents a stationary phase for the minimum of $H_{11} ^1-H_{22} ^1$
coinciding with the minimum of the squared mass difference $\tilde
m_2 ^2-\tilde m_1 ^2$, and it can be used to calculate the level crossing
probability \cite{GBA93} around the lower resonance. To obtain the correct
stationary
phase to analyse the higher resonance, we rotate the evolution matrix given
in Eq. (\ref{Eq.1}) in the following way:
\begin{equation}
\label{Eq.35}H=e ^{-i\psi \Lambda _7}he ^{i\psi \Lambda _7}
\end{equation}
Now the minimum of the  difference $H_{33}-H_{11}$ indicates a stationary
phase which now coincides with the required resonance.

\section{Conclusions}

Resonances represent a crucial region in the time evolution of neutrinos
oscillating in matter. They are closely related with the nonadiabatic
character of the oscillation. We investigated a general criterion to define
a resonant region when three neutrino are present in the oscillation
phenomenon. We observed that two of the three commonly employed criteria to
identify a resonance in two neutrino oscillations are not valid anymore in
its simplest extension to three neutrino MSW effect. For instance, mixing
angles can present maximal values far from resonant regions and therefore this
criterion to define a resonance has to be abandoned. Furthermore, stationary
phases do not necessarily coincides with resonant regions. The safest way to
 identify such resonance regions is to investigate the behavior of the squared
matter eigenvalue
differences, looking for their minimum values.

\section{Acknowledgements}

The authors would like to thank Funda\c c\~ao de Amparo \`a Pesquisa do
Estado de S\~ao Paulo (FAPESP), Con\-se\-lho Na\-cio\-nal de
De\-sen\-vol\-vi\-men\-to Cien\-t\'\i \-fi\-co e Tec\-no\-l\'o\-gi\-co
(CNPq) and Coordena\c c\~ao de Aperfei\c coamento de Pessoal de N\'\i vel
Superior (CAPES) for several financial supports.

\newpage

\centerline{\bf CAPTIONS}

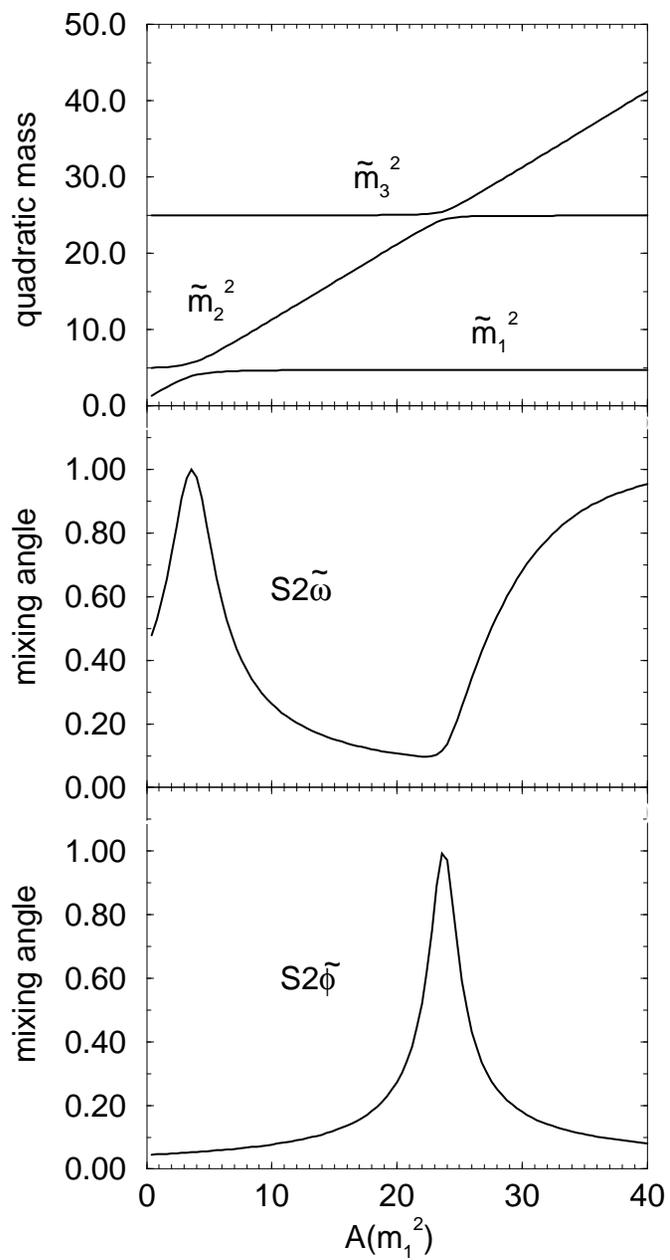
\vskip.5cm \noindent {\bf Figure 1:} Squared matter eingenvalues
$\tilde m_i ^2$ and the relevant matter mixing angles $\tilde \omega $
and $\tilde \phi $ as a function of the parameter $A$ are presented.
The values of the parameters $\phi ,\omega , $ and $m_i ^2$ in vacuum:
$m_3 ^2=5m_2 ^2=25m_1 ^2;$ $\sin ^2\phi =5\times 10 ^{-4};$ $\sin
^2\omega \cos ^2\phi =5\times 10 ^{-2}$  were chosen in order to well
demonstrate the behavior of these parameters as a function of $A$. The
eingenvalues $\tilde m_i$ and the quantity $A$ are given in units of
$m_1 ^2$.

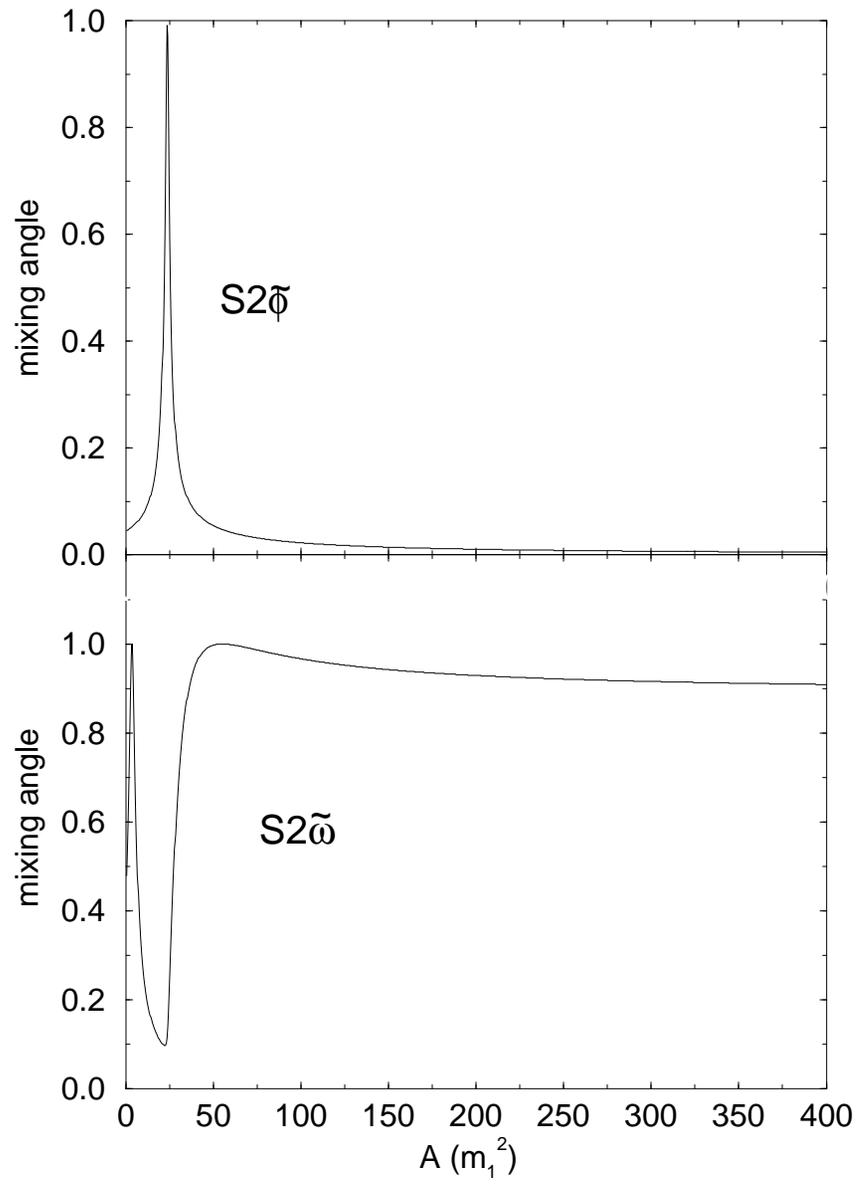
\vskip.5cm \noindent {\bf Figure 2:} Mixing angles $\tilde \omega$ and
$\tilde\phi$ as a function of the $A$ for larger values of $A$. The
parameters $\phi ,\omega ,$ and $m_i ^2$ are the same as that of the
Figure 1.

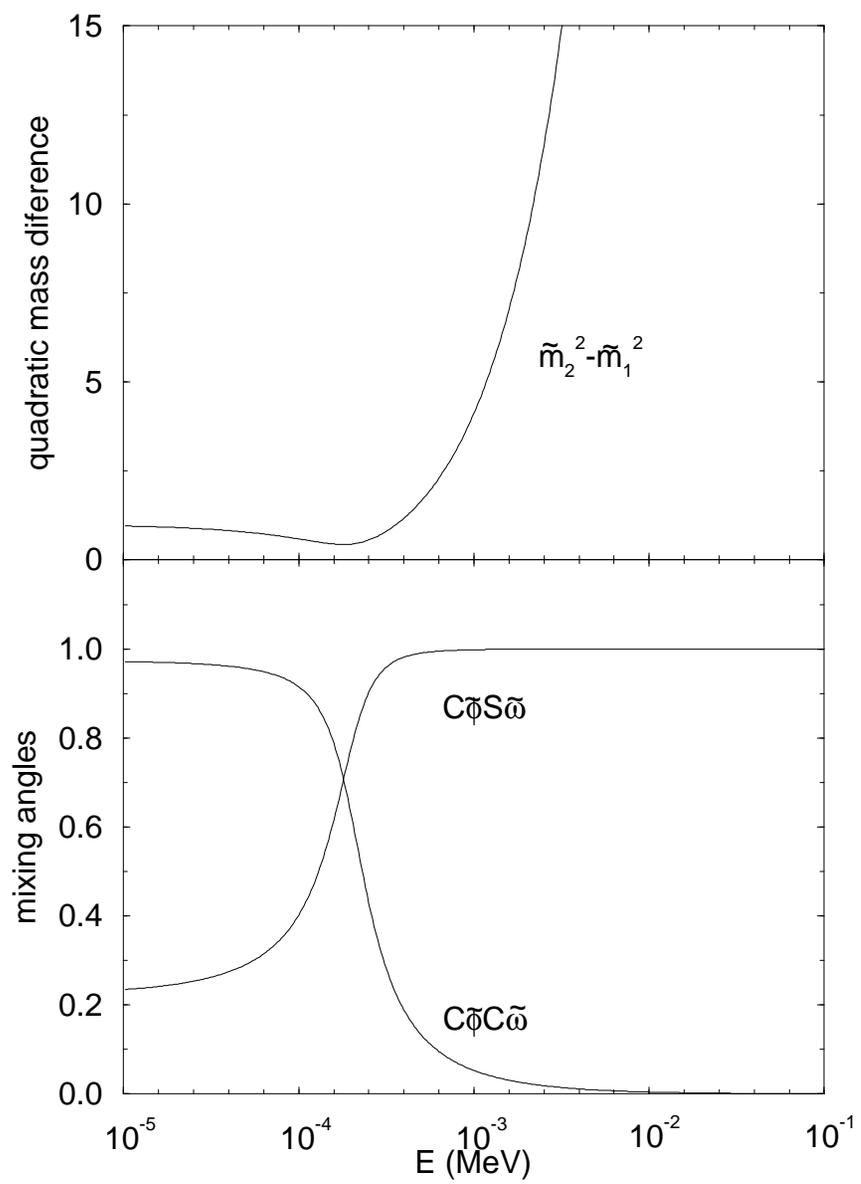
\vskip.5cm \noindent {\bf Figure 3:} The squared mass difference
$\tilde m_2 ^2-\tilde m_1 ^2$ , and the quantities $\cos \tilde \phi
\sin \tilde \omega $ and $\cos \tilde \phi \cos \tilde \omega $
quantities are presented as a function of the energy of neutrinos. The
values of the parameters $\phi ,\omega $ and $m_i ^2$ in vacuum are:
$m_3 ^2=1.445\times 10 ^{-4}eV ^2;$ $m_2 ^2=10 ^{-8}$eV$^2;$ $m_1
^2=0;$ $\sin ^2\phi =5\times 10 ^{-4};$ $\sin ^2\omega = 0.050025.$ The
squared masses  are given in units of $m_2 ^2.$

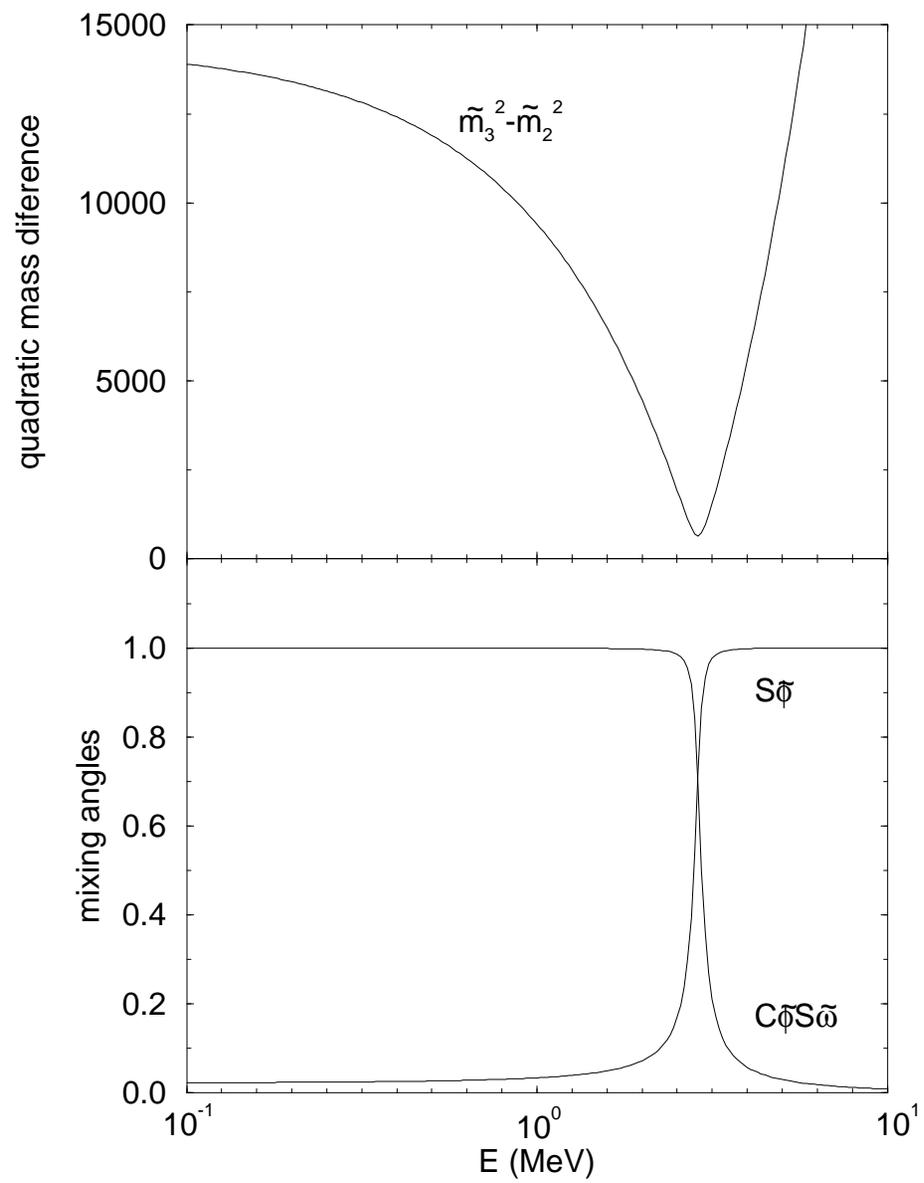
\vskip.5cm \noindent {\bf Figure 4:} The squared mass difference
$\tilde m_3 ^2-\tilde m_2 ^2$ , the effective $\cos \tilde \phi \sin
\tilde \omega $ and sin$\tilde \phi $ quantities are presented as a
function of the energy of neutrinos. The values of the parameters $\phi$
, $\omega $ and $m_i^2$ in vacuum are the same ones used to draw Fig. 3.

\end{document}